# Nitrogen-tailored quasiparticle energy gaps of polyynes

Kan Zhang (张侃)[1], Jiling Li (李继玲)[1], Peitao Liu (刘培涛)[2], Guowei Yang (杨国伟)[1], Lei Shi (石磊)[1*]

[1] *State Key Laboratory of Optoelectronic Materials and Technologies, Nanotechnology Research Center, Guangzhou Key Laboratory of Flexible Electronic Materials and Wearable Devices, School of Materials Science and Engineering, Sun Yat-sen University, Guangzhou 510275, P. R. China*

[2] *Shenyang National Laboratory for Materials Science, Institute of Metal Research, Chinese Academy of Sciences, 110016 Shenyang, China*

* Correspondence: shilei26@mail.sysu.edu.cn

Polyyne, a sp$^1$-hybridized linear allotrope of carbon, has a tunable quasiparticle energy gap, which depends on the terminated chemical ending groups as well as the chain length. Previously, nitrogen doping was utilized to tailor the properties of different kinds of allotrope of carbon. However, how the nitrogen doping tailors the properties of the polyyne remains unexplored. Here, we applied the GW method to study the quasiparticle energy gaps of the N-doped polyynes with different lengths. When a C atom is substituted by a N atom in a polyyne, the quasiparticle energy gap varies with the substituted position in the polyyne. The modification is particularly pronounced when the second-nearest-neighboring carbon atom of a hydrogen atom is substituted. In addition, the nitrogen doping makes the Fermi level closer to the lowest unoccupied molecular orbital, resulting in a n-type semiconductor. Our results suggest another route to tailor the electronic properties of polyyne in addition to the length of polyyne and the terminated chemical ending groups.

## 1. Introduction

Polyyne composed of alternative single-triple bonds has been extensively studied on its conductivities as a wide-gap semiconductor varied with the length,[1-5] ending chemical groups,[6-8] and atomic structures (linear, bent, and cyclo).[9-13] The conjugated triple bonds cause delocalization of valence electrons filling into π-orbitals, thus introducing energy gap near the Fermi level.[14] The energy gap is determined by the value of bond length alternation (BLA) of the polyyne, which is the bond length difference between single and triple bonds because of Peierls distortion.[15] Experimental and theoretical studies suggested that both the length of a polyyne and the ending groups enable to change the BLA, thus influence the properties of the polyyne.

Usually, substitution doping with heteroatoms is often used to regulate the electronic property of carbon materials. For example, heteroatoms including B and N have been introduced into carbon nanotubes or graphene to modify both the conduction band minimum and valence band maximum, resulting in improved electronic properties.[16-23] Isoelectronic doping was proposed to modify the energy gap of graphdiyne, a material with sp- and $sp^2$ hybridization.[24,25] However, how the substitution doping influences the properties of the polyyne remains unexplored.

It is well established that calculations based on Hartree-Fock functional overestimate the quasiparticle energy gap of the polyyne,[26] while density functional theory (DFT) calculations underestimate the quasiparticle energy gap.[27] In contrast, the GW method[28-32] (G and W represent the Green's function and the screened Coulomb interaction, respectively) provides a good approximation for the self-energy of a many-body system of electrons, and therefore, has been proved as a reliable method for obtaining more accurate quasiparticle energies for a variety of systems.[33-37] For the polyyne, previously, two methods, i.e., the B3LYP hybrid functional based on Coulomb-attenuating method (CAM) and the diffusion quantum Monte Carlo method (DMC), have been acknowledged to predict results of quasiparticle energy. Thus, an energy gap of 6.52 eV for a hydrogen-capped polyyne with 10 carbon atoms and an energy gap of 3.61 eV for a carbyne (an infinitely long polyyne) were obtained by employing CAM-B3LYP[27] and DMC[37], respectively. Here, we applied the GW method to calculate the quasiparticle energy gaps of the polyyne and the carbyne. The obtained gaps for hydrogen-capped polyyne with 10 carbon atoms and carbyne are 6.75 and 3.53 eV, respectively, which are consistent with the predicted values.[27, 37] Then we applied the GW method to predict the quasiparticle energy gaps of N-doped polyynes considering different substituted positions of the N atom into a polyyne with different length. Compared to the pristine polyyne, the quasiparticle energy gap strongly varies with the substituted position. Moreover, the polyyne changes into a n-type semiconductor with the nitrogen substitution. Our work provides an efficient route to tune the electronic properties of polyyne and especially substituting the second-nearest-neighboring carbon atom allows to achieve the lowest energy gap close to visible region, which could be applicable in applications of optical devices in future.

2. **Computational method**

Quasiparticle energy gaps of pristine and N-doped polyynes were calculated by the GW method. The structural models were constructed for polyyne with different length, as shown Figure 1. To ensure only one molecule in one cell and no cross-linking between molecules, a H-caped (undoped and N-doped) polyyne molecule was placed along the z-axis (OC, O as the original point) in a hexagonal cell. The relaxation was performed by using the PBEsol functional,[38] the interaction between ions and electrons was described by the projector-augmented wave (PAW) method[39] as implemented in the Vienna ab-initio simulation package (VASP)[40, 41] and the cut-off energy was set to 450 eV. Note that for polyyne molecule, a sufficient vacuum length of 10 Å was used to keep the molecule an isolated object excluding any spurious image interactions. DFT electronic structure calculations were carried out by using Quantum ESPRESSO.[42, 43] Only gamma point was used. Quasiparticle energies were calculated by using the BerkeleyGW code.[32, 44] Then by using the Wannier90 suit,[45-47] the projections of C/N s and p orbitals were calculated and wannier interpolated quasiparticle band structures including the quasiparticle energy gap were finally calculated (HOMO-LUMO gap for polyyne molecule and energy gap for polyyne molecule-stacked bulk materials).

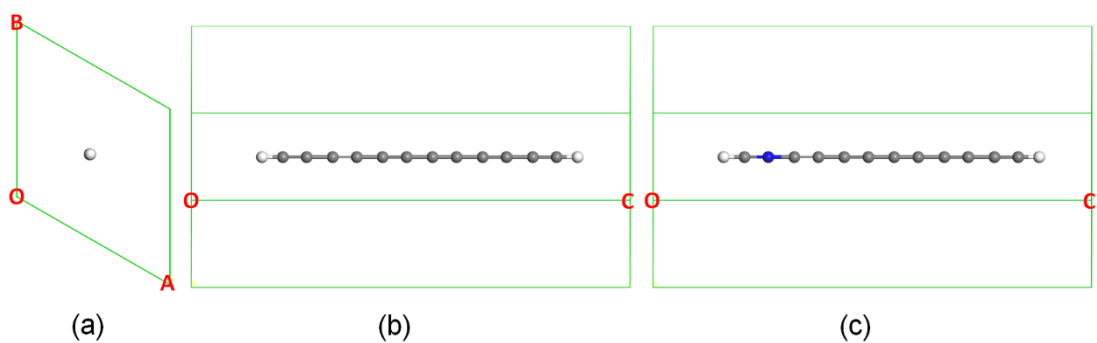

(a)　　　　　　　(b)　　　　　　　(c)

Figure 1 Computational models in (a) top view, (b) side view of primitive cell of polyyne[12] and (c) C2→N-polyyne[12] where OA, OB and OC correspond to the lengths of lattice parameters a, b and c, respectively. C, H and N atoms are in gray, white and blue, respectively.

## 3. Results and discussion

Since the value of BLA determines the HOMO-LUMO gap of polyyne, we calculated the bond lengths of single and triple bonds for polyynes containing 10-20 carbon atoms (labeled as polyyne[n], where n is the number of carbon atoms in a polyyne) and corresponding N-doped polyynes[n] at different positions (labeled as Cm→N-polyyne[n], where m is the counting number of the carbon atom except hydrogen atom from the end of a polyyne as illustrated in Figure 1). As shown in Figure 2a (full results can be found in Figure S1), we can clearly see that the bond lengths of both single and triple bonds are not evenly distributed in a polyyne. Generally, the bond length is shorter in the middle of the polyyne. Also, when the polyyne is longer, the bond length difference between the bond at the end and the bond in the middle gets larger, suggesting a length-dependent property. In addition, the average BLA increases as the length of the polyyne decreases, because the p electron delocalizes from σ bond to π bond, resulting in a larger electronic energy gap.[14] The introduction of nitrogen atom in a polyyne greatly modulates the length of the carbon-carbon bond close to the nitrogen (Figure 2b), thus modifying the properties of the polyyne. Similar results were found when the nitrogen atom locates at different positions of a polyyne (Figure 2c). Therefore, it is with great interest to see how the nitrogen at different position affects the quasiparticle energy gap.

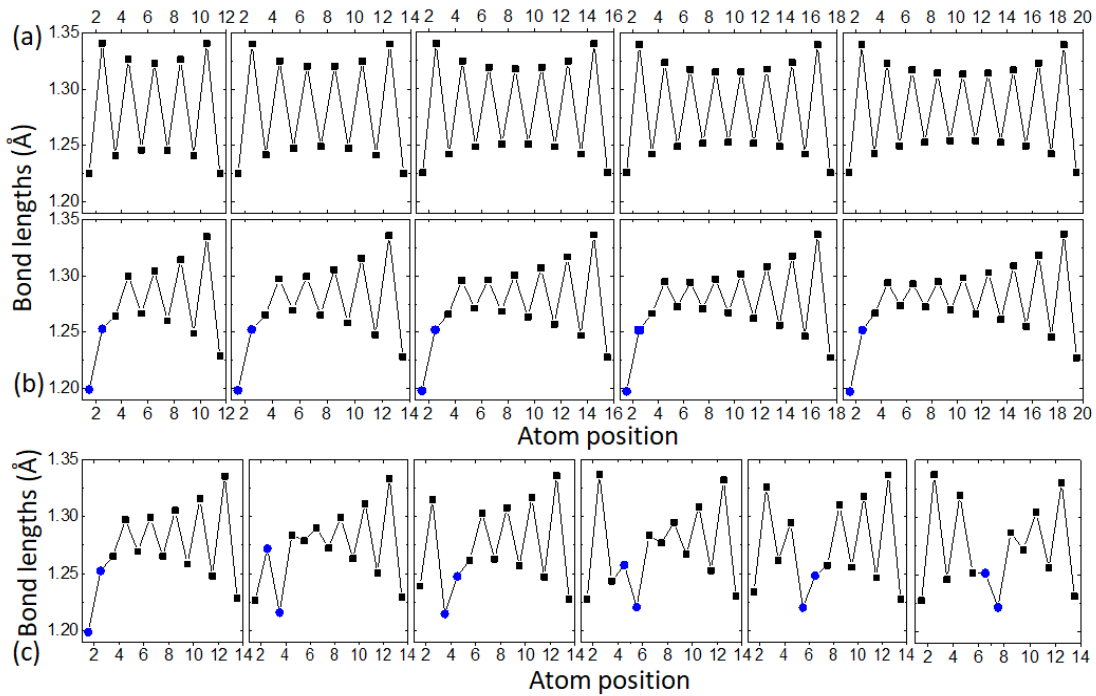

Figure 2 Bond lengths of (a) polyyne[n], (b) C2→N-polyyne[n], and (c) Cm→N-polyyne[14]. n=12,14,16,18, 20 from left to right and blue dots show lengths of C-N bonds.

The orbitals were calculated by using the obtained bond lengths of the polyynes. As shown in Figure 3, from the orbitals of polyyne[14], it is clearly seen that a conjugated π bond system exists through the chain. The orbitals of single and triple bonds are orthogonal. With nitrogen doping, for the C2→N-polyyne[14] the orbitals are greatly modified. The doping position in the polyyne determines how much the orbitals can be altered, indicating a N-position dependent property. The same is found for the C2→N-polyyne[12].

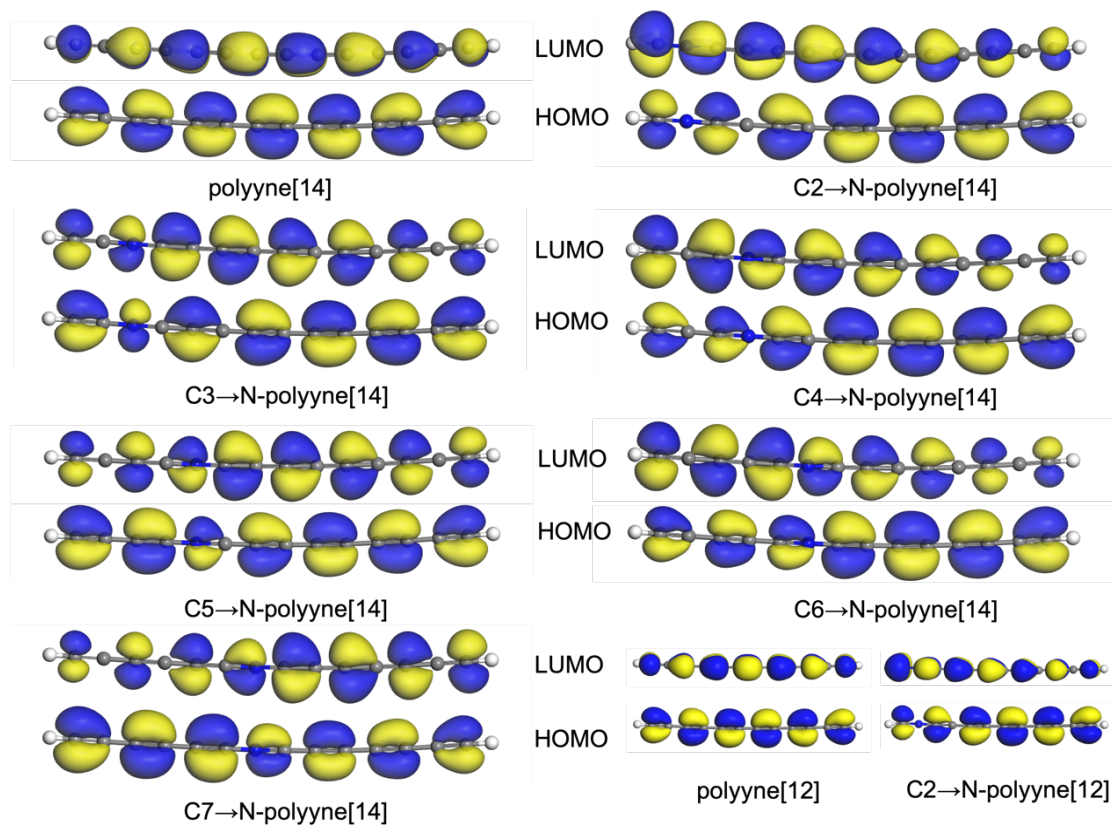

Figure 3 Orbitals of polyyne[14] and Cm→N-polyyne[14]. Orbitals of polyyne[12] and C2→N-polyyne[12] (bottom right). Positive and negative orbital wave functions are marked in blue and yellow, respectively.

Our calculated quasiparticle energy gaps of polyyne[n] confirm the length-dependent property (Figures 4 and 5), which is consistent with previous results.[48-50] The LUMOs of the polyyne[n] only slightly move down when the polyyne gets longer, whereas the HOMO is much closer to the Fermi level (Figure 4a), resulting in a reduced HOMO-LUMO gap. Compared to the undoped polyyne, when doped with nitrogen at the second position (C2→N-polyyne[n]), two main obvious consequences have been observed. One is that the HOMO-LUMO gap decreases remarkably, manifested by the downshifted LUMO and the upshifted HOMO (Figures 4 and 5a). The other is that the Fermi level is getting close to the downshifted LUMO (according to Figures 5b and 5c, more p electron states are provided due to the N-doping), presenting a behavior of n-type semiconductor.[51, 52] In contrast, substitutions at the other positions only slightly move the LUMO down. Therefore, the conductivity of polyyne can be improved by the nitrogen doping, and the substitution position is crucial to obtain the optimal effect. This reveals that the experimental research could focus more on the synthesis of C2→N-polyyne[n] among all the Cm→N-polyyne[n]. However, since the C2→N-polyyne[14] have the lowest LUMO and highest HOMO, suggesting the most instability of the molecule, which makes it more challenging to be synthesized.

The HOMO-LUMO gaps of polyyne[n] and corresponding Cm→N-polyyne[n], n=12-20, are summarized in Figure 5. Again, the HOMO-LUMO gap changes greatly with the nitrogen position in a polyyne. Especially when the second carbon is substituted with a nitrogen, i.e., the C2→N-polyyne[n], the reduction of the HOMO-LUMO gaps is more pronounced. When the n is large enough, the HOMO-LUMO gap of C2→N-polyyne[n] tends to saturate at a value of below 2 eV, which is close to the value obtained for the carbyne inside carbon nanotubes [49,53]. In addition, when the doping position moves from the end to the middle of the polyyne, the gap tends to saturate at a constant value, which can even be greater than that of undoped polyyne (Figure 5a). Our findings not only suggest a new strategy to engineer the quasiparticle energy gap of a polyyne via doping the polyyne with nitrogen at selective positions, but also reveal a way to increase the electric conductivity via nitrogen doping like most of the carbon materials, [54, 55] which would benefit its application as a semiconductor.

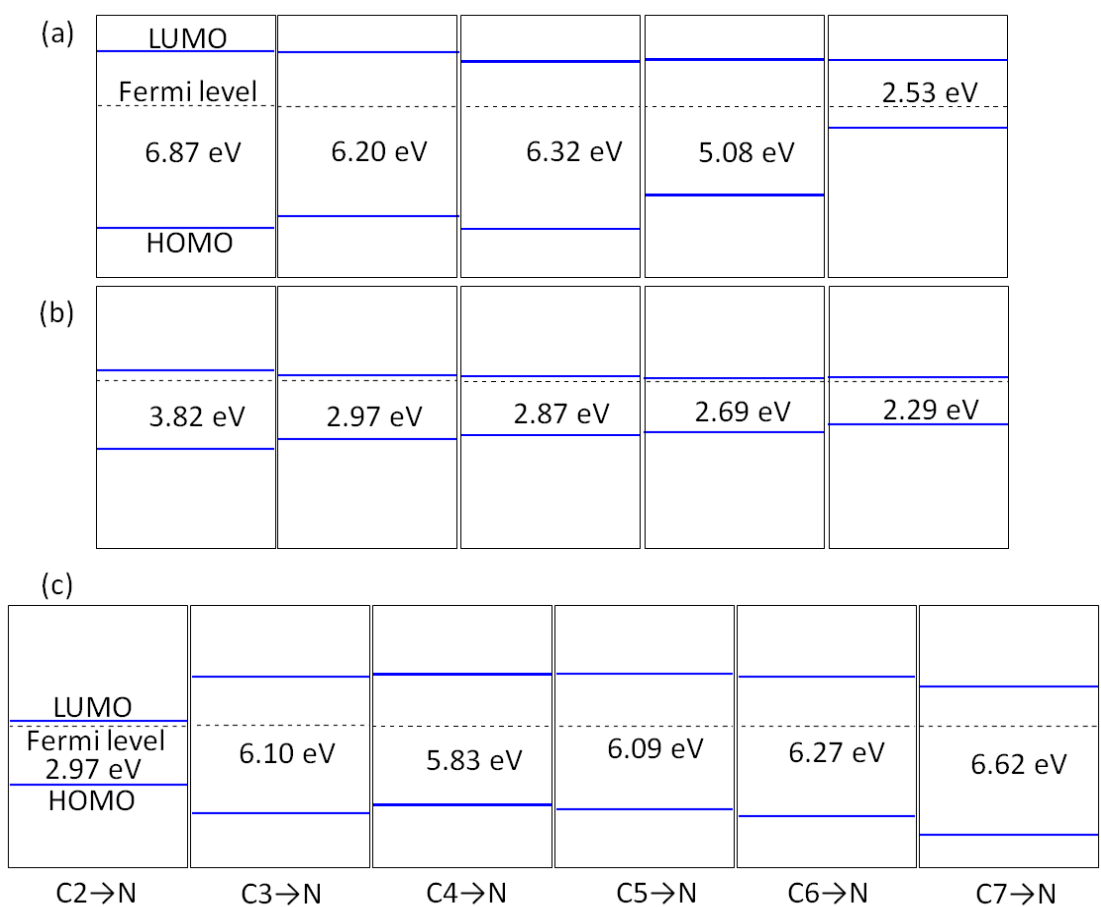

Figure 4 HOMOs and LUMOs of (a) polyyne[n], (b) C2→N-polyyne[n] (n=12,14,16,18, 20 from left to right), and (c) Cm→N-polyyne[14].

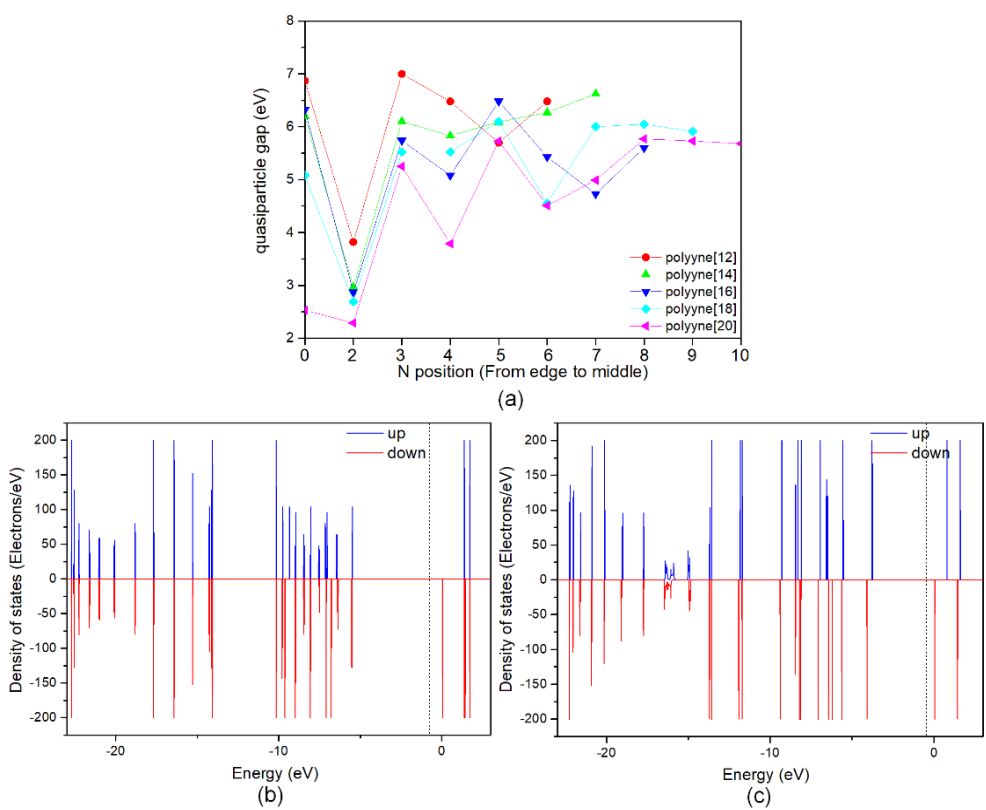

Figure 5 (a) HOMO-LUMO gaps of polyynes[n] and Cm→N-polyyne[n]. PDOS of p orbitals of (b) undoped polyyne[12] (c) C2→N-polyyne[12].

Polyynes are usually synthesized and kept in solvent due to its particular synthesis process and instability. Recently, stacked polyynes in a form of small crystal has been prepared, which suggests that the stacked polyynes can be stable after drying.[56] When the polyyne molecules are stacked into bulk materials, it is also expected that the properties of the stacked polyynes would change, as inspired from the changed property of two-dimensional materials compared to that of their bulk three-dimensional counterparts, e.g., graphene versus graphite. [57, 58] To view the difference on electronic properties between stacked polyynes and single polyyne molecule, the interactions among molecules should be considered in the calculation. Our calculations show that the quasiparticle energy gap of stacked polyynes decreases as compared to that of the polyyne molecules (Figure S2), suggesting another modulated way to further reduce the energy gap of a polyyne except nitrogen doping, which requires specific study in future.

4. **Conclusion**

In summary, we theoretically studied how the nitrogen substitution doping tailors the quasiparticle energy gaps of the polyyne molecules with different lengths. We find that the N-doping enables to greatly change the gap of the polyyne and the gap highly depends on the doping position in the polyyne. Substituting the second C atom

(counting from the end of a polyyne except hydrogen atom) by N atom reduces the gap most due to the significantly decreased BLA, whereas the N-doping in the middle of a polyyne only slightly varies the gap. Our study reveals an effective route to tune the electronic properties of polyyne via N-doping, which would benefit the future applications of the polyyne as a n-type semiconductor. In addition, we suggest that stacking the polyyne molecule could also tailor the quasiparticle energy gap, which deserves further investigations in future.

See Supplementary Material for the complete bond lengths and quasiparticle energy gaps of single non-doped/N-doped polyynes.

**Acknowledgment**
We thank Dr. Zhuhua Zhang for helpful discussion. This work was supported by Guangdong Basic and Applied Basic Research Foundation (2019A1515011227), National Natural Science Foundation of China (51902353), Fundamental Research Funds for the Central Universities, Sun Yat-sen University (22lgqb03) and State Key Laboratory of Optoelectronic Materials and Technologies (OEMT-2022-ZRC-01).